\documentclass[conference]{IEEEtran}
\IEEEoverridecommandlockouts
\usepackage{cite}
\usepackage{amsmath,amssymb,amsfonts}
\usepackage{algorithmic}
\usepackage{graphicx}
\usepackage{textcomp}
\usepackage{xcolor}
\def\mB{\mbox{$\mathbf{B}$}}
\def\mD{\mbox{$\mathbf{D}$}}

\def\mH{\mbox{$\mathbf{H}$}}

\def\mL{\mbox{$\mathbf{L}$}}

\def\mU{\mbox{$\mathbf{U}$}}

\def\mz{\mbox{$\mathbf{z}$}}

\def\mx{\mbox{$\mathbf{x}$}}

\def\ms{\mbox{$\mathbf{s}$}}
\def\my{\mbox{$\mathbf{y}$}}

\newcommand{\beq}{\begin{equation}}
\newcommand{\eeq}{\end{equation}}
\def\BibTeX{{\rm B\kern-.05em{\sc i\kern-.025em b}\kern-.08em
    T\kern-.1667em\lower.7ex\hbox{E}\kern-.125emX}}
\begin{document}

\title{Robust Filter Design for  Graph Signals\\
{\footnotesize}
\thanks{This work was partially supported  by  the MIUR under the PRIN Liquid-Edge contract,  by the  Huawei Technology France SASU, under agreement N. TC20220919044 and by the European Union under the Italian National Recovery and Resilience Plan (NRRP) of NextGenerationEU, partnership on “Telecommunications of the Future” (PE00000001 - program “RESTART”).}
}

\author{Lucia Testa, Stefania Sardellitti,  and Sergio Barbarossa\\
Sapienza Univ. of Rome, DIET Dept, Via Eudossiana 18, 00184, Rome, Italy\\ e-mail: \{lucia.testa, stefania.sardellitti, sergio.barbarossa\}@uniroma1.it}
\maketitle




\begin{abstract}
Our goal  in this paper is the robust design of filters  acting on  signals observed  over graphs subject to  small perturbations of their edges. 
The focus is on developing a method to identify spectral and polynomial graph filters that can adapt to the perturbations in the underlying graph structure while ensuring the filters adhere to the desired spectral mask. To address this, we propose a novel approach that leverages approximate closed-form expressions for the perturbed eigendecomposition of the Laplacian matrix associated with the nominal topology. Furthermore, when dealing with noisy input signals for graph filters, we propose a strategy for designing FIR filters  that jointly minimize the approximation error with respect to  the ideal filter and the estimation error of the output, ensuring robustness against both graph perturbations and noise. Numerical results validate the effectiveness of our proposed strategies, highlighting  their capability to efficiently manage perturbations and noise.
\end{abstract}

\begin{IEEEkeywords}
Graph perturbation, robust graph filters, graph signal processing. 
\end{IEEEkeywords}

\section{Introduction}

Graph Signal Processing (GSP) \cite{ortega2018graph}, \cite{Shuman2013} has  recently emerged as a powerful framework providing tools for the analysis and processing of data defined over graphs.
Graph-based representations are pivotal tools for extracting information from data across various fields, ranging from finance, communication and social networks to biological sciences. 
While in certain contexts, such as physical networks, the graph topology might be perfectly known,  in many others, the topology is completely unknown and has to be inferred from the observed data. Furthermore, there are situations where our knowledge of the graph topology is not perfect  but affected by random uncertainties.
For instance, in wireless communication networks, the topology is known, but some links may inadvertently drop due to random blocking or fading\cite{Ghosh2007}. In such cases, we may only assume to know a \text{nominal} graph, whose topology may be perturbed by some random edge dropping.
Similarly, in brain networks, the interaction among different regions of the brain changes over time \cite{Calhoun2014},\cite{Preti2017}, and in biological networks, temporal variations of the network topology describing protein–protein and protein–DNA interactions are observed \cite{Kim2014}.
In data-driven networks, the topology is inferred from the data, and the  learning accuracy depends on  the inference algorithm   as well as on the observed data that may be corrupted by noise or outliers.
Therefore, it becomes interesting to analyze uncertain graphs, i.e., graphs wherein some edges may be altered  with a certain probability.

Assuming  that only a small percentage of edges remains uncertain,  this paper aims  to study the impact of small perturbations  on the design of robust spectral and finite impulse response (FIR) filters acting on signals defined over graphs. 
Graph Filters (GFs) have been extensively studied  in the literature  \cite{Moura_14}, \cite{Segarra_17},\cite{LiuFilter}. Similarly,
the stability  of graph  filters to perturbations has been thoroughly investigated in previous works \cite{Moura_14}, \cite{Gama2020},\cite{LiuFilter}
,\cite{Isufi_22}, \cite{Marques_23}.
A preliminary study on the impact of perturbations of  graphs and simplicial complexes on the robustness of filters acting on  signals observed  over  such domains is discussed in 
\cite{sardellitti2022}.\\
Recently, in \cite{Marques_23} the authors introduced a novel approach that jointly addresses robust graph filter identification and graph denoising. In \cite{kenlay2021}
the authors studied the stability of spectral GFs when a small number of edge were rewired. In \cite{ceci2020graph} structural equation models are combined with total least squares to jointly infer signals and perturbations. Recently, the stability of Graph Convolutional Neural Networks (GCNs) has attracted increasing interest and has been the subject of extensive investigation in various works, e.g. \cite{gama},\cite{Keriven},\cite{Levie}, \cite{testa2023stability}.\\
Our focus in this work  is to design  robust  graph filters acting on signals defined over graphs that closely approximate the desired filter on the nominal graph, despite minor edge alterations.
 The proposed approach leverages the small perturbation analysis of the graph Laplacian eigenpairs developed in \cite{Ceci_Barb}. We use first-order closed form expressions for the perturbed Laplacian matrix eigenvalues/eigenvectors pairs to design graph  filters that are  robust against topology uncertainties. Then,  we  express robust filters in closed form, which  depends only on the known probabilities of edge perturbations.
 Finally, in case where the filer input signals are affected by random noise, we introduce an optimization strategy aimed at finding a FIR filter that exhibits robustness both to graph perturbations and to noise interference. 
 Specifically, the optimal FIR filter is designed to minimize, jointly, the approximation error respect to the  desired (unperturbed) filter and the 
estimation error in the filter output. The effectiveness of the proposed strategies is substantiated through numerical results, demonstrating their good  performance in handling both perturbations and noise.

\section{Small Perturbation Analysis of Graph Laplacian}
In this section, first
 we  quickly review some of the key tools for processing signals defined over graphs. Thereafter,
we briefly recall the theory of small perturbation analysis of the graph Laplacian eigendecomposition developed in \cite{Ceci_Barb}. Let us consider an undirected graph $\mathcal{G} = \{\mathcal{V}, \mathcal{E}\}$ 
composed of a set  $\mathcal{V}=\{1,\ldots, N\}$ of $N$  nodes and  a set $\mathcal{E}$ of  edges, with cardinality $|\mathcal{E}|=E$.
The  connectivity of the graph can be described through the incidence matrix $\mB_1 \in \mathbb{N \times E}$, whose columns $\{\mathbf{b}_m\}_{m=1}^E$ establish which nodes are incident to each edge $m$.  Specifically, given an arbitrary orientation of the edges, the  entries of the column vector $\mathbf{b}_m$ are all zero except  the entries $\mathbf{b}_m(i_s)=1$ and $\mathbf{b}_m(i_t)=-1$ corresponding to the indices $i_s$ and $i_t$ of the endpoints of edge $m$.  
The graph Laplacian $\mathbf{L} \in \mathbb{R}^{N \times N}$, is a symmetric, semidefinite positive  matrix that captures the connectivity property of the graph and is defined as $\mL=\mB_1 \mB_1^T= \sum_{i \in \mathcal{E}} \mathbf{b}_i \mathbf{b}_i^T$.
We assume, w.l.o.g., that the graph is connected and we denote with $\mL=\mU \boldsymbol{\Lambda} \mU^T$ the Laplacian eigendecomposition, where $\mU$ is the matrix whose columns are the eigenvectors $\mathbf{u}_i$, $i=1,\ldots,N$ and $\boldsymbol{\Lambda}$ is the diagonal matrix with entries the associated eiegenvalues $\lambda_i$, $i=1,\ldots,N$.
We assume that the eigenvalues are listed in increasing order.\\
A signal $\mathbf{s}$ on a graph $\mathcal{G}$ is defined as a mapping from the vertex set
to the set of real numbers, i.e. $\mathbf{s}: \mathcal{V}\rightarrow \mathbb{R}$.
For undirected graphs, the GFT $\hat{\mathbf{s}}$ of a graph signal $\mathbf{s}$ is defined as the projection of $\mathbf{s}$ onto the subspace spanned by the eigenvectors $\mU=\{\mathbf{u}_i\}_{i=1}^{N}$ of the Laplacian matrix, see e.g. \cite{Shuman2013}, \cite{Pesenson2008}, i.e.  $\hat{\mathbf{s}}=\mU^T \mathbf{s}$.\\
\textbf{Graph Small Perturbation Analysis.}
A small perturbation of the graph $\mathcal{G}$ corresponds to add or remove a few edges, thus  resulting in the perturbed graph $\tilde{\mathcal{G}}$. Let us denote with $\mathcal{E}_p \subset \mathcal{E}$ the set of  edges of the nominal graph that are altered.  The perturbed graph is  described by the  perturbed  Laplacian matrix $\widetilde{\mathbf{L}} = \mathbf{L} + \Delta\mathbf{L}$, where  $\Delta\mathbf{L} \in \mathbb{R}^{N \times N}$ is a perturbation matrix that can be expressed as 
\begin{equation}
    \Delta\mathbf{L} = \sum_{m \in \mathcal{E}_p} \sigma_m \mathbf{b}_m\mathbf{b}_m^T,
\end{equation}
where   $\sigma_m=1$ if edge $m$ is added and $\sigma_m=-1$, if edge $m$ is removed. 
Clearly, the perturbation of the nominal Laplacian $\mL$ induces  a perturbation of its eigenvalues decomposition denoted by
\beq
\widetilde{\mathbf{L}} = \mathbf{L} + \Delta\mathbf{L}=\widetilde{\mU} \widetilde{\boldsymbol{\Lambda}}\widetilde{\mU}^T
\eeq
where $\widetilde{\mU}:=\mU+\Delta\mathbf{U}$ and 
$\widetilde{\boldsymbol{\Lambda}}:=\boldsymbol{\Lambda}+\Delta\boldsymbol{\Lambda}$
denote the eigenvector and the eigenvalues matrices of $\widetilde{\mathbf{L}}$.
In the case where all eigenvalues are distinct
and the perturbation affects a few percentage of edges, the perturbed eigenvectors $\{\widetilde{\mathbf{u}}_i\}_{i=1}^N$ and eigenvalues $\{\widetilde{\lambda}_i\}_{i=1}^N$ of $\widetilde{\mathbf{L}}$ are related to the unperturbed eigenvectors $\{\mathbf{u}_i\}_{i=1}^N$ and eigenvalues $\{\lambda_i\}_{i=1}^N$ of $\mathbf{L}$, by the following approximations \cite{Ceci_Barb}:

\label{eq:pert_eigenvalues}
\begin{align}
    &\widetilde{\lambda}_i \approx \lambda_i+\delta \lambda_i=\lambda_i + \mathbf{u}_i^T\Delta\mathbf{L}\mathbf{u}_i 
\end{align}

    \beq
    \label{eq:pert_eigenvectors}
     \widetilde{\mathbf{u}}_i \approx \mathbf{u}_i+\delta \mathbf{u}_i=\mathbf{u}_i + \sum_{j \neq i} \frac{\mathbf{u}_j^T\Delta\mathbf{L}\mathbf{u}_i}{\lambda_i - \lambda_j} \mathbf{u}_j.
\eeq
Therefore, from (\ref{eq:pert_eigenvalues}) and (\ref{eq:pert_eigenvectors}), the perturbations $\delta\lambda_i$ and $\delta\mathbf{u}_i$ of the $i$-th eigenvalue and eigenvector, respectively, can be written as \cite{Ceci_Barb}:
\begin{equation}
\label{eq:variationlambda}
    \begin{split}
    & \delta\lambda_i = \sum_{m \in \mathcal{E}_p} \sigma_m\mathbf{u}_i^T\mathbf{b}_m\mathbf{b}_m^T\mathbf{u}_i 
    \end{split}
\end{equation}
\begin{equation}
\label{eq:variation}
  \begin{split}
    & \delta\mathbf{u}_i = \sum_{m \in \mathcal{E}_p} \sigma_m \delta \mathbf{u}_{i m}
\end{split}
\end{equation}
with $\delta \mathbf{u}_{i m}= \sum_{j \neq i} \frac{\mathbf{u}_j^T\mathbf{b}_m\mathbf{b}_m^T\mathbf{u}_i}{\lambda_i - \lambda_j} \mathbf{u}_j$.
The above formulas come from a first-order perturbation analysis \cite{Ceci_Barb} which effectively  captures the impact of the perturbation on the graph topology. In particular,  the 
term $\mathbf{u}_i^T \mathbf{b}_m \mathbf{b}_m^T \mathbf{u}_i= (u_i(v_{s,m})-u_i(v_{t,m}))^2$ in (\ref{eq:variation}) is a measure of the variation of the eigenvector $\mathbf{u}_i$ at the vertices $v_{s,m}$ and $v_{t,m}$ of edge $m$. Then, the largest perturbations occur over the  edges that exhibit the highest eigenvectors variation. For example, in graphs characterized by dense clusters, the edges with most significant perturbations are the inter-cluster edges. Additionally, we can observe that the eigenvector corresponding to
the null eigenvalue does not affect any other eigenvalue/eigenvector, and eigenvectors associated with eigenvalues very close to
each other typically experience large perturbations.
\section{Robust Spectral Filtering over Perturbed Graphs}
Our goal in this section is to design  robust spectral filters for graphs that undergo small perturbation of their edges. In general, a spectral filter operating over a graph signal $\mathbf{s}_0$ can be modelled as 
\beq
\mathbf{y}_0=\mH \ms_0
\eeq
where $\mH=\mU \mD \mU^T$
and  $\mD$ is a diagonal matrix with entries the spectral mask coefficients $\{h(\lambda_i)\}_{i=1}^{N}$  that we wish to implement over the eigenvalues of $\mL$. Let us now assume  that the graph topology is altered by either the addition or removal of a few edges. 
Then, using the closed-form  expression  in 
 (\ref{eq:pert_eigenvectors}), we introduce  the matrix $\tilde{\mU}$ whose columns are the eigenvectors $\tilde{\mathbf{u}}_i$ of the perturbed Laplacian $\tilde{\mL}$. \\
Our objective is to design a filter, denoted as $\tilde{\mH}_{\mathbf{\tilde{D}}}=\tilde{\mU} \tilde{\mD} \tilde{\mU}^T$,   approximating with  minimum averaged error the desired spectral mask $\mD$, while being robust against small perturbation in the graph topology,  such as the addition or removal of edges. 
  Assuming that the  perturbation of an edge $m$,  is a random event characterized  by a certain probability $p_m$, 
  we describe this perturbation  as a binary r.v. $Z_m$ equal to $1$, with probability $p_m$, if edge $m$ is perturbed, and $0$ otherwise.  
  Then, 
let us consider the   filter estimation error averaged with respect to random edge perturbations, i.e.
\begin{equation}
\label{eq:spectmaskFormula}
\begin{aligned}
    & \mathbb{E}\left[\left\| \mathbf{U} \mathbf{D} \mathbf{U}^T - \tilde{\mathbf{U}}\tilde{\mathbf{D}}\tilde{\mathbf{U}}^T \right\|^2_{F}\right] = \mathbb{E}\left[\left \| \tilde{\mathbf{R}} - \tilde{\mathbf{D}}\right\|^2_{F}\right]
\end{aligned}
\end{equation}
where $\tilde{\mathbf{R}}=\tilde{\mathbf{U}}^T\mathbf{U} \mathbf{D} \mathbf{U}^T\tilde{\mathbf{U}}$ and  the last equality follows from the unitary property of $\tilde{\mathbf{U}}^T$.

Our goal is to derive the optimal spectral coefficients as solution of the following optimization problem
\begin{equation}
\label{eq:pr_1}
\begin{split}
\min_{\tilde{\mathbf{D}}\in \mathbb{R}^{N \times N}} &  \; f(\mathbf{\tilde{D}}):=\mathbb{E}\left[\left \| \tilde{\mathbf{R}} - \tilde{\mathbf{D}}\right\|^2_{F}\right]\\
 \text{s.t.} & \quad \tilde{d}_{ij}=0, \quad \forall i\neq j 
\end{split}    
\end{equation}
where the constraints force the matrix $\tilde{\mD}$ to be diagonal.
Problem in (\ref{eq:pr_1}) is convex and its 
 optimal solution $\tilde{\mathbf{D}}^{\star}$ can be easily derived in closed-form by averaging with respect to the random edge perturbations. Specifically, the optimal \(\tilde{\mathbf{D}}^{\star}\) is derived by taking the expectation of the diagonal elements of $\tilde{\mathbf{R}}$, 
i.e.
\begin{align}    \mathbf{\tilde{D}}_{ii}^{\star} = \mathbb{E}[\tilde{\mathbf{R}}_{ii} ], \quad i=1,\ldots,N.
\end{align}
To derive closed-form expressions for these optimal diagonal entries, let us  recall that \(\tilde{\mathbf{R}}=\tilde{\mathbf{U}}^T\mathbf{U} \mathbf{D} \mathbf{U}^T\tilde{\mathbf{U}}\). Then, using the equality $\mH=\mU \mD \mU^T$, we get  
$\tilde{\mathbf{R}}=\mathbf{\tilde{U}}\mathbf{H}\mathbf{\tilde{U}^T}$ and it holds
\beq
\label{eq:E_1}
\tilde{\mathbf{D}}_{ii}^{\star}=\mathbb{E}[\tilde{\mathbf{u}}^T_{i}\mathbf{H}\tilde{\mathbf{u}}_{i}]. 
\eeq
Assuming the random variables $Z_m$ and $Z_n$, i.i.d. for $m\neq n$, we get
$\mathbb{E}[Z_m Z_n] = E[Z_m^2] = p_m$, if $m=n$ and   $\mathbb{E}[Z_m Z_n] = p_m p_n$ if $m \neq n$.
Then, using (\ref{eq:pert_eigenvectors}) and taking in (\ref{eq:E_1}) the expectation values with respect to the random variables $Z_m$,
we  easily derive the optimal solution 
\begin{equation}
\label{eq:optmask}
\begin{aligned}
    &\tilde{\mathbf{D}}_{ii}^{\star}=    \mathbf{u}_{i}^{T}\mathbf{H}\mathbf{u}_{i}+\!    \sum_{k,l=1}^{N}\mathbf{H}_{k,l}\left( \sum_{m \in \mathcal{E}_p} p_m \delta \mathbf{u}_{i,m} (k)\delta \mathbf{u}_{i,m} (l) \right.\\ &\left.+\sum_{m \in \mathcal{E}_p}\sum_{n \in \mathcal{E}_p} p_m p_n \sigma_m \sigma_n\delta \mathbf{u}_{i,m}(k)\delta \mathbf{u}_{i,n}(l)\right)\\
       &+2 \mathbf{u}_{i}^{T}\mathbf{H}\left(\sum_{m \in \mathcal{E}_p} p_{m} \sigma_m \delta \mathbf{u}_{i,m}\right).
\end{aligned}
\end{equation}
Note that the optimal spectral mask  consists of  a first term that represents the optimal solution for an unperturbed graph while the other terms  take into account  the perturbation statistics, enabling  the mask to be adapted to random changes in the graph's topology.
\section{Robust Localized Filters}
A desired graph filter $\mH$  may be efficiently parameterized using a polynomial FIR filter of order $L$ based on the graph Laplacian as  \cite{Moura_14}, \cite{Segarra_17}
\beq
\label{eq:H_FIR}
\mH=\sum_{k=0}^{L} h_k \mL^k.
\eeq
This is a local operator that combines graph signals from neighbors  of each node, at a $k$-hop distance, through the coefficients $\{h_k\}_{k=0}^{L}$.  
Assuming that the graph undergoes random edge perturbations, our goal in this section is 
to design a robust FIR filter 
approximating with  minimum averaged error the desired FIR filter $\mH$ while being robust to small perturbations of the graph topology.
Then, our aim is to derive the optimal filter $\tilde{\mH}=\sum_{k=0}^{L} \tilde{h}_k \tilde{\mL}^{k}$ with minimum average distance from the nominal filter $\mH$ in  (\ref{eq:H_FIR}).
Hence, we may formulate the following optimization problem  
\begin{equation}
\label{eq:prob_2}
\min_{\tilde{\mathbf{h}} \in \mathbb{R}^{L+1}} 
 \; g(\tilde{\mathbf{h}}):= \mathbb{E}\left[\left\| \sum_{k=0}^L \tilde{h}_k \tilde{\mathbf{L}}^k - \sum_{n=0}^L h_n\mathbf{L}^n\right\|^2_{F}\right] 
\end{equation}
where the expectation is taken with respect to the random edge perturbations.
Note that the objective function can be easily written as
\begin{equation}
\begin{split}
 g(\tilde{\mathbf{h}})=& \mathbb{E}\left[\left\| \sum_{k=0}^L \tilde{h}_k \tilde{\mathbf{U}}\tilde{\mathbf{\Lambda}}^k\tilde{\mathbf{U}}^T - \sum_{n=0}^L h_n\mathbf{U}\mathbf{\Lambda}^n\mathbf{U}^T \right\|^2_{F}\right]  \\   = & \mathbb{E}\left[ \left\| \sum_{k=0}^L \tilde{h}_k \tilde{\mathbf{\Lambda}}^k - \mathbf{M} \right\|^2_{F}\right],\\
\end{split}
\end{equation}
where \(\mathbf{M} = \tilde{\mathbf{U}}^T\mathbf{U}\left(\sum_{n=0}^L h_n\mathbf{\Lambda}^n\right)\mathbf{U}^T\tilde{\mathbf{U}}\). Then, defining the vector $\mathbf{m}=\text{diag}(\mathbf{M})$, whose entries are the diagonal elements of $\mathbf{M}$,    the problem in (\ref{eq:prob_2}) is equivalent to the following one
\begin{equation}
\label{eq:pr_h2}
\begin{aligned}
\min_{\tilde{\mathbf{h}} \in \mathbb{R}^{L+1}} \; \mathbb{E}[\| \mathbf{m} - \mathbf{\tilde{\Phi}}\tilde{\mathbf{h}} \|^2_{F}]
\end{aligned}
\end{equation}
with $\mathbf{\tilde{\Phi}}=[\mathbf{1}, \tilde{\boldsymbol{\lambda}},\ldots,\tilde{\boldsymbol{\lambda}}^{L} ]$, $\tilde{\boldsymbol{\lambda}}^k=\{\tilde{\lambda}_i^k\}_{i=1}^N$, $k=1,\ldots,L$.

It can be easily shown that the optimal solution $\tilde{\mathbf{h}}^{\star}$ of the problem in (\ref{eq:pr_h2}) is given by \beq \tilde{\mathbf{h}}^{\star}=  \mathbb{E}[\mathbf{\tilde{\Phi}}^T\mathbf{\tilde{\Phi}}]^{-1}\mathbb{E}[\mathbf{\tilde{\Phi}}^T\mathbf{m}].\eeq
Let us now derive a closed form expression for  $\tilde{\mathbf{h}}^{\star}$ by taking the expectation with respect to the random edge perturbations. Note that it holds:
\begin{equation}
    \begin{aligned}
        \mathbf{\tilde{\Phi}}^T\mathbf{\tilde{\Phi}} = 
        \begin{bmatrix}
    N &  \sum_{i=1}^{N} \tilde{\lambda}_{i} &\cdots  & \sum_{i=1}^{N} \tilde{\lambda}_{i}^{L} \\
    \vdots & \ddots & \vdots \\
    \sum_{i=1}^{N}\tilde{\lambda_i}^L & \sum_{i=1}^{N}\tilde{\lambda_i}^{L+1} &\cdots  & \sum_{i=1}^{N} \tilde{\lambda}_{i}^{2 L}
\end{bmatrix}.
    \end{aligned}
\end{equation}
From (\ref{eq:variationlambda})  we have $\delta \lambda_{i} = \sum_{m \in \mathcal{E}_p} Z_m  q_{i,m}$ with $q_{i,m}=\mathbf{u}_i^T\mathbf{b}_m\mathbf{b}_m^T\mathbf{u}_i$. Then, we get
the following expectation values 

\begin{equation}
\label{eq:Edeltalambda}
    \begin{aligned}
&\mathbb{E}\left[\sum_{i=1}^{N}\tilde{\lambda}_{i}\right] = \sum_{i=1}^{N}(\lambda_{i} + \sum_{m \in \mathcal{E}_p} p_m \delta \lambda_{i,m}) \medskip\\
    &\mathbb{E}\left[ \sum_{i=1}^{N}\tilde{\lambda}_{i}^{k}\right] 
   = \sum_{i=1}^{N}\sum_{j=0}^{k} \binom{k}{j}\lambda_{i}^{k-j}\mathbb{E}\left[(\delta \lambda_{i})^j\right]    
\end{aligned}
\end{equation}
where the term $\mathbb{E}\left[(\delta \lambda_{i})^j\right] = \mathbb{E}\left[(\sum_{m \in \mathcal{E}_p} Z_m  q_{i,m})^{j} \right]$
can be easily calculated under the assumption of i.i.d. variables 
by using the equalities $\mathbb{E}[Z_m^j] = p_m^j$, $\forall m \in \mathcal{E}_p$ and $\mathbb{E}[Z_r Z_s \ldots Z_j] = p_r p_s\ldots p_j$ for any set of distinct indexes $r,s,\ldots,j$ within $\mathcal{E}_p$.
Finally, let us consider the vector $\mathbb{E}[\mathbf{\tilde{\Phi}}^T\mathbf{m}]$ whose entries can be expressed as:
\begin{equation}
    \label{eq:Phi_m}
    \begin{aligned}
        \mathbb{E}[(\mathbf{\tilde{\Phi}}^T\mathbf{m})_{i}]&= \left(\sum_{n=0}^L h_n \lambda_i^n\right)\left( \sum_{p=1}^N \mathbf{u}_i(p)^4 \mathbb{E}[(\lambda_i + \delta \lambda_i)^L] \right. \\  & + \mathbf{u}_i(p)^2
        \mathbb{E}[\delta \mathbf{u}_i(p)^2(\lambda_i + \delta \lambda_i)^L] \\  &+ 
        \left.  2\mathbf{u}_i(p)^3\mathbb{E}[\delta \mathbf{u}_i(p)(\lambda_i + \delta \lambda_i)^L] \right).
    \end{aligned}
\end{equation}

The first term of this sum can be simplified as in (\ref{eq:Edeltalambda}), the second term can be written as $\mathbb{E}[\delta \mathbf{u}_{i}(p)^2]\mathbb{E}[(\lambda_i + \delta \lambda_i)^L]$, where $\mathbb{E}[\delta \mathbf{u}_{i}^2(p)]=
\sum_{m \in \mathcal{E}_p}\sum_{n \in \mathcal{E}_p}\sigma_m \sigma_n\delta \mathbf{u}_{i,m} (k)\delta\mathbf{u}_{i,n}(l)p_m p_n$. Finally, we can write the third term as $2\mathbf{u}_i(p)^3\mathbb{E}[\delta \mathbf{u}_i(p)]\mathbb{E}[(\lambda_i + \delta \lambda_i)^L]$, with $\mathbb{E}[\delta \mathbf{u}_{i}(p)]=\sum_{m \in \mathcal{E}_p} p_{m} \sigma_m \delta \mathbf{u}_{i,m}(p)$.

\section{Robust filtering for noisy signals and graph perturbations}
\label{sec:noisy signal}
In this section we focus on the robust filtering of noisy graph signals. Our goal is to find the optimal FIR filter which is robust against both graph perturbations and signal noise. Let us assume that the filter's input graph signal $\mx$ is  affected by random noise $\mathbf{n}$ and define $\my=\mx+\mathbf{n}$ as  the noisy signal we aim to filter. Then, in the case where the graph is perturbed and the signal is affected by noise, the  filtered signal is given by $\tilde{\mz}=\tilde{\mH} \my$. Then,  we are interested in finding optimal coefficients that minimize jointly
 the
approximation error respect to the desired (unperturbed) filter $\mH$, i.e. $g(\tilde{\mathbf{h}})$ in (\ref{eq:prob_2}),
and the estimation error in the perturbed filter output with respect to the ideal output $\mH \mx$.
Hence, we may formulated the following optimization problem:
\begin{equation}
\label{eq:pr_h5}
\begin{split}
\min_{\tilde{\mathbf{h}} \in \mathbb{R}^{L+1}} \; & \mathbb{E}\left[\left\| \sum_{k=0}^L \tilde{h}_k \tilde{\mathbf{L}}^k - \sum_{n=0}^L h_n\mathbf{L}^n\right\|^2_{F}\right]\\ & + \gamma \, \mathbb{E}\left[\left\| \sum_{k=0}^L \tilde{h}_k \tilde{\mathbf{L}}^k \my - \sum_{n=0}^L h_n\mathbf{L}^n \mx \right\|^2_{F}\right]
\end{split}
\end{equation}
where $\gamma$ is  a non-negative coefficient to control the trade-off between the two estimation errors.
Note that the optimization problem in (\ref{eq:pr_h5}) can be written in the following equivalent form  
\begin{equation}
\label{eq:pr_h32}
\begin{aligned}
\min_{\tilde{\mathbf{h}} \in \mathbb{R}^{L+1}} \; \mathcal{L}(\tilde{\mathbf{h}}):=\mathbb{E}[\| \mathbf{m} - \mathbf{\tilde{\Phi}}\tilde{\mathbf{h}} \|^2_{F}]+ \gamma \, \mathbb{E}[\| {\mathbf{w}} - \mD_{y}\mathbf{\tilde{\Phi}}\tilde{\mathbf{h}} \|^2_{F}]
\end{aligned}
\end{equation}
where $\mathbf{w}=\tilde{\mU}^T \mU (\sum_{n=0}^L h_n \boldsymbol{\Lambda}^n) \mU^T\mathbf{x}$  and $\mD_{y}=\text{diag}(\hat{\mathbf{y}})$ 
 with $\mathbf{\hat{y}}= \mathbf{\tilde{U}}^T\mathbf{y}$.
 The objective function $ \mathcal{L}(\tilde{\mathbf{h}})$ of the  problem   in (\ref{eq:pr_h32}) can be expressed as 
\begin{equation}
    \begin{aligned}
        \mathcal{L}(\tilde{\mathbf{h}}) &= \mathbb{E} \Bigl[\text{tr}\Bigl\{\mathbf{m}^T\mathbf{m} + \mathbf{\tilde{h}}^T\mathbf{\tilde{\Phi}}^T\mathbf{\tilde{\Phi}}\mathbf{\tilde{h}} -2\mathbf{\tilde{h}}^T\mathbf{\tilde{\Phi}}^T\mathbf{m}\Bigr\}\Bigr]+\\
        &\gamma \mathbb{E}\Bigl[\text{tr} \Bigl\{ \mathbf{w}^T\mathbf{w} + \mathbf{\tilde{h}}^T\mathbf{\tilde{\Phi}}^T\mathbf{D}_{y}^2\mathbf{\tilde{\Phi}}\mathbf{\tilde{h}} -2\mathbf{{\tilde{h}}}^T\mathbf{\tilde{\Phi}}^T\mathbf{D}_{y}\mathbf{w}\Bigr\} \Bigr]
    \end{aligned}
\end{equation}
from which the optimal solution $\mathbf{\tilde{h}}$ can be derived in closed form as  
\begin{equation}
    \begin{aligned}
        \mathbf{\tilde{h}}= \mathbb{E}\Bigl[\mathbf{\tilde{\Phi}}^T\mathbf{\tilde{\Phi}} + \gamma\mathbf{\tilde{\Phi}}^T{\mathbf{D}}_{y}^2\mathbf{\tilde{\Phi}}\Bigr]^{-1}\mathbb{E}\Bigl[\mathbf{\tilde{\Phi}}^T(\mathbf{m}+\gamma \mathbf{D}_{y}\mathbf{w})\Bigr].
    \end{aligned}
\end{equation}
Following similar derivations as in (\ref{eq:Edeltalambda}) and (\ref{eq:Phi_m}), closed form solutions can be obtained, which  we omit here due to space constraints.




\section{Numerical Results}
\begin{figure}
\centering
\includegraphics[width=\columnwidth,height=5.5cm]{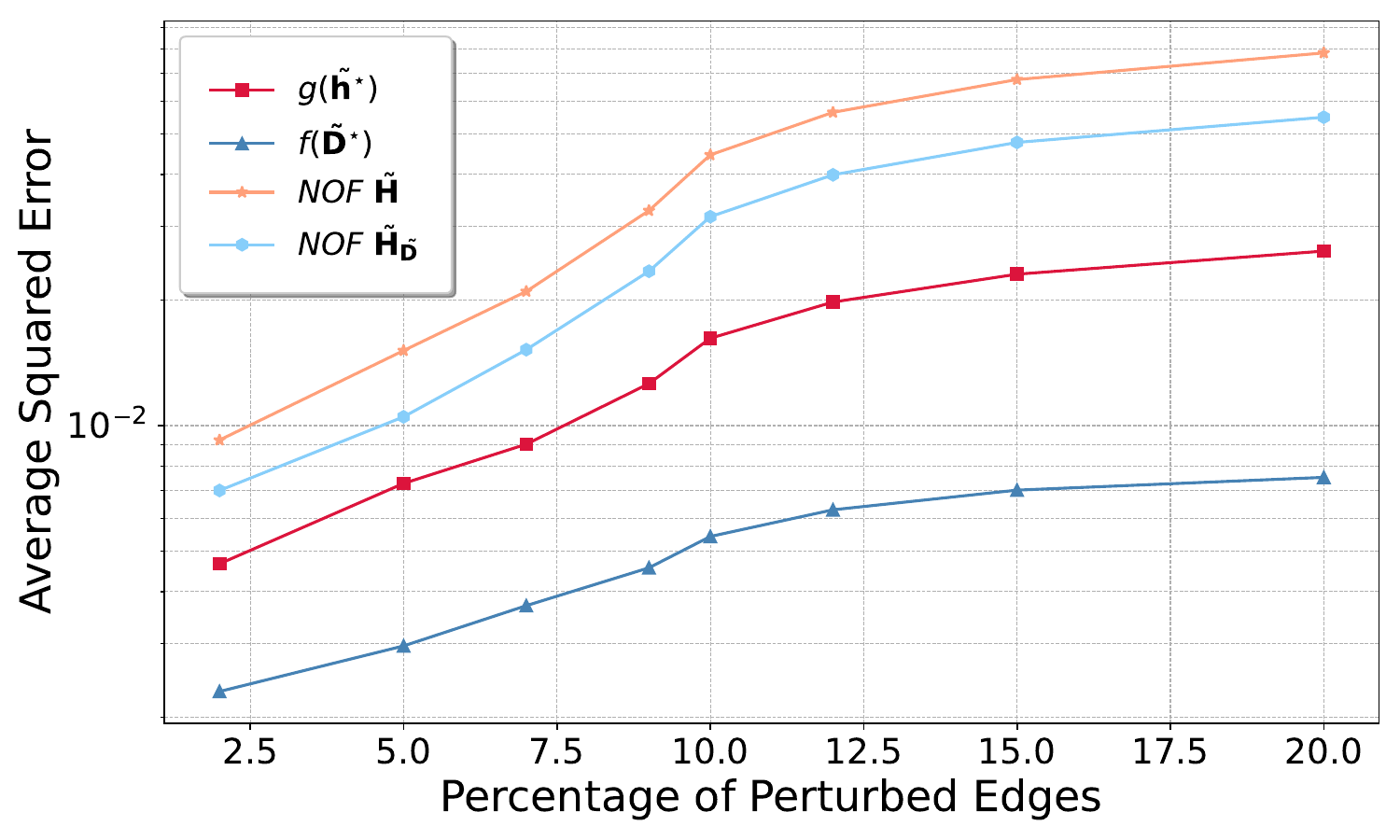} %
\caption{Average squared errors vs the percentage of perturbed edges.} \label{fig:percpert}
\end{figure}

\begin{figure}
\centering
\includegraphics[width=\columnwidth,height=5.5cm]{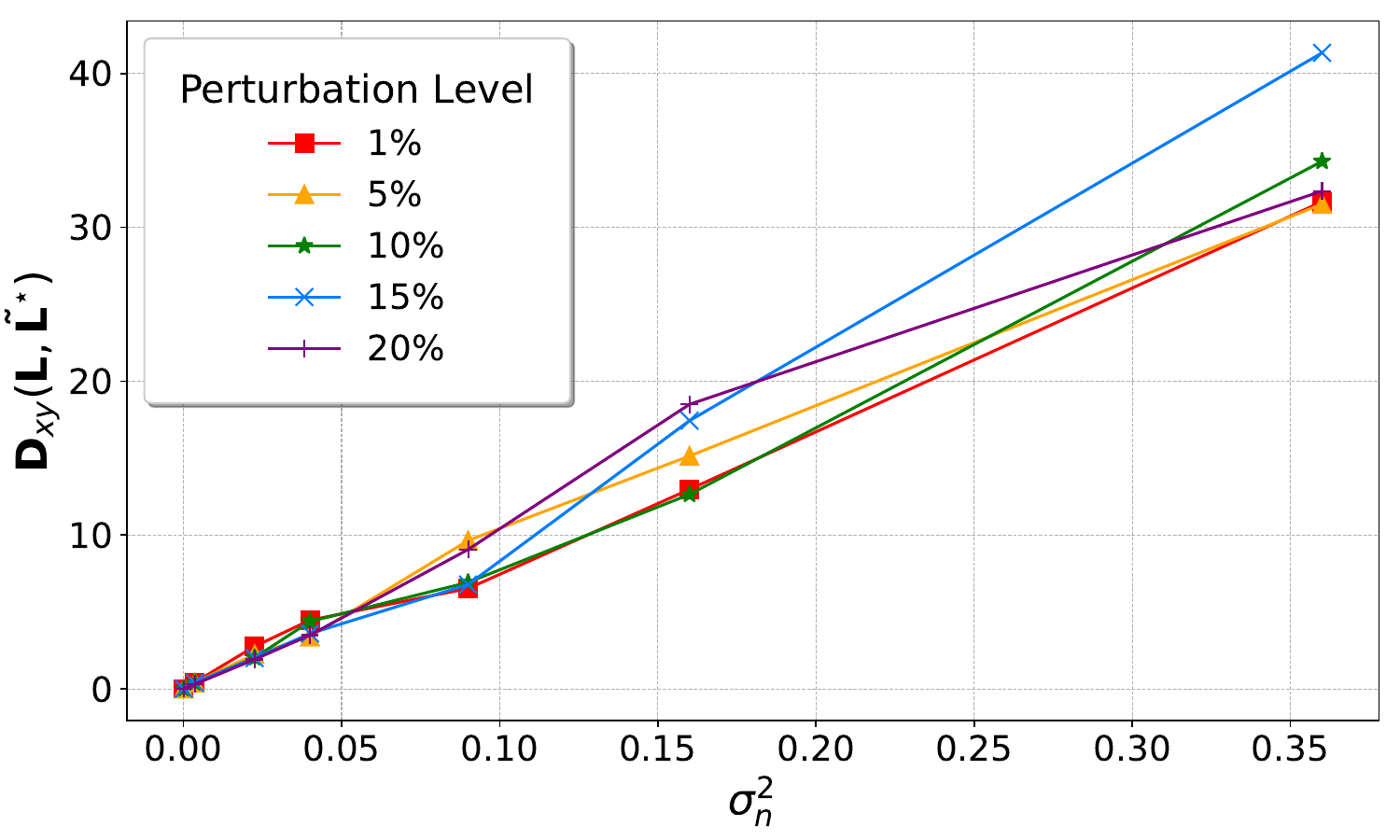} 
\caption{Average distance  $\mathbf{D}_{x y}(\mathbf{L},\mathbf{\tilde{L}}^{\star})$ vs   the noise variance $\sigma_n^2$. }\label{figure:SNR}
\end{figure}

\begin{figure}
\centering
\includegraphics[width=\columnwidth,height=5.5cm]{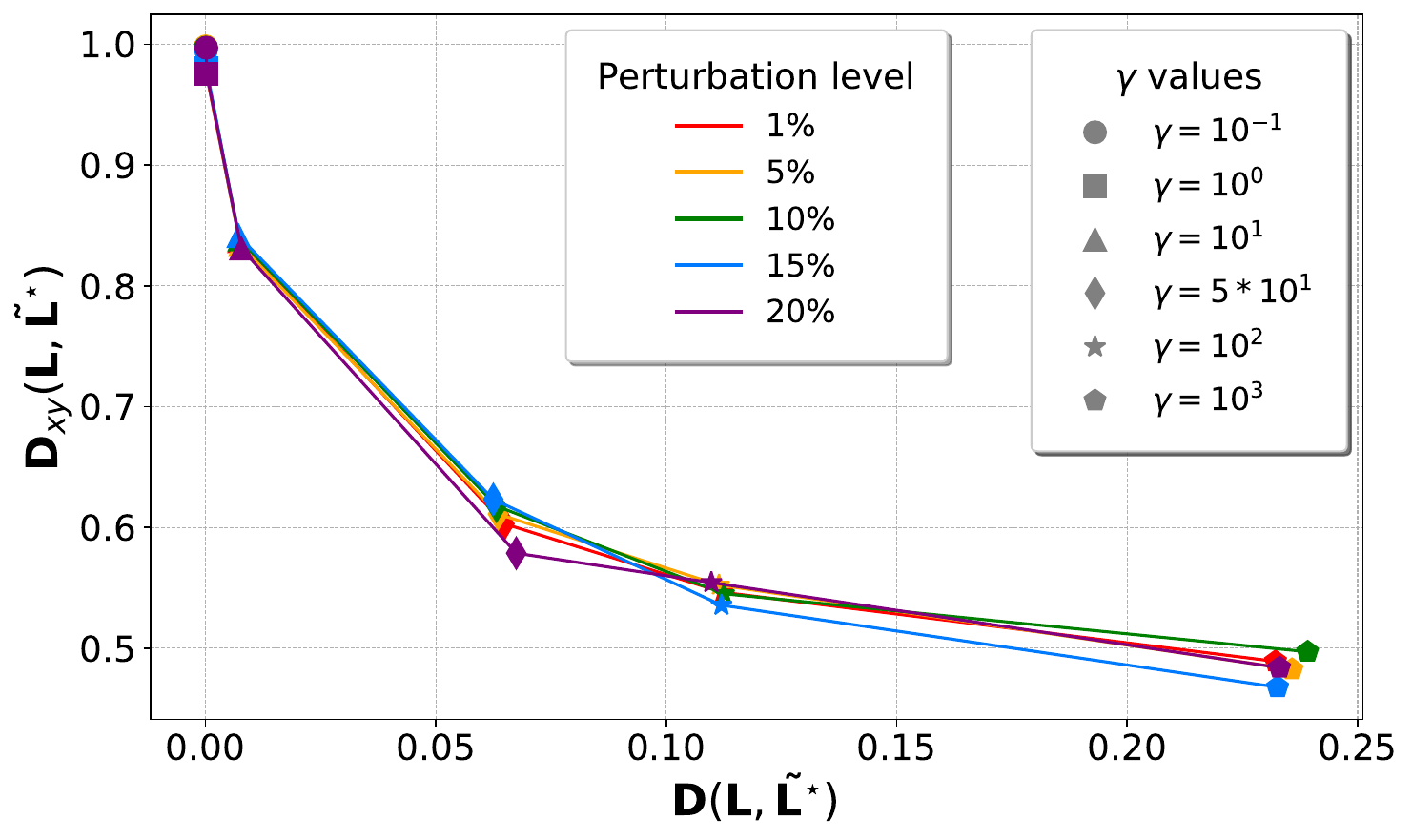} 
\caption{Trade-off between $\mathbf{D}(\mathbf{L},\mathbf{\tilde{L}}^{\star})$ and $\mathbf{D}_{xy}(\mathbf{L},\mathbf{\tilde{L}}^{\star})$ varying $\gamma$.}\label{fig:gamma}
\end{figure}

In this section, we validate the effectiveness of the proposed filtering designs through comprehensive numerical experiments.\\
To evaluate the robustness of both  spectral and FIR  filters against perturbed graphs, we generated $100$  random realizations of graphs, each consisting of two clusters.
These graphs have an intra-cluster and inter-cluster connection probability equal to $0.7$ and $0.08$, respectively. We randomly    perturbed a varying percentage of the total edges, from $1\%$  to $20\%$, ensuring  that after  perturbations the graph remains connected. Using 
the closed-form 
formulas in 
(\ref{eq:optmask}), 
 (\ref{eq:Edeltalambda}) and (\ref{eq:Phi_m}), 
we derived, respectively, the optimal mask $\tilde{\mathbf{D}}^{\star}$ and the optimal FIR coefficients $\tilde{\mathbf{h}}^{\star}$.
Then, we calculated for both filters the associated average  distances between the desired and estimated filters, i.e. $f(\tilde{\mD}^{\star})$ and $g(\tilde{\mathbf{h}}^{\star})$, respectively.
In Fig. \ref{fig:percpert}, we report these average distances versus the percentage of perturbed edges.
Furthermore, to assess the robustness of our proposed optimal filters,  we also report in Fig. \ref{fig:percpert} the average distance between the ideal target filter and the non-optimized filters  (NOF) where the filter coefficients are directly derived using the perturbed graph,  for both the spectral and FIR filters. It is evident  the robustness enhancement achieved  using   the proposed optimal filtering design.
 However,  the spectral filtering approach exhibits more robustness,  due to the fact that the FIR filter is a polynomial approximation of  the intended one.\\ 
\noindent
Let us now consider the robust filtering of noisy signals. 
In order to generalize our findings, in this experiment we employed Erd\H{o}s--R\'enyi graphs. We generated $100$ different graphs, each containing  $100$ nodes, with a connection probability of $p=0.5$ for each edge, ensuring that the graphs  remain connected after the perturbations. 
Then, by solving the problem in (\ref{eq:pr_h5}), 
let us denote with 
$\mathbf{D}(\mathbf{L},\mathbf{\tilde{L}}^{\star})$
and
$\mathbf{D}_{x y}(\mathbf{L},\mathbf{\tilde{L}}^{\star})$ 
the optimal (normalized) values of the first and second term of the objective function.
In Fig. \ref{figure:SNR}  we report the optimal distance $\mathbf{D}_{x y}(\mathbf{L},\mathbf{\tilde{L}}^{\star})$,  representing the estimation error in the perturbed filter output, versus the noise variance $\sigma_n^2$ and for various levels of  perturbation.
We can observe as increasing  the variance of the noise, $\mathbf{D}_{x y}$ increases as well, albeit it maintains robustness for small perturbation levels.\\
Finally,  
 in Fig. \ref{fig:gamma} we illustrate the trade-off between the two terms, $\mathbf{D}_{x y}(\mathbf{L},\mathbf{\tilde{L}}^{\star})$ and $\mathbf{D}(\mathbf{L},\mathbf{\tilde{L}}^{\star})$, by varying the penalty coefficient $\gamma$. It can be observed as the optimal filter approximation error (respect to the desired filter) and the  optimal  estimation error in the perturbed filter output exhibit  robustness across various percentages of  edges perturbations.


\section{Conclusions}
In this paper we consider the robust  design of filters for signals observed over graphs that may undergo random small perturbations of their topology.  Our objective is to devise a strategy  for deriving spectral and polynomial filters that can  adapt to small changes in the graph's topology, while still  closely approximating  the desired spectral mask. To this end, we introduce an innovative approach that utilizes approximate closed-form solutions for the perturbed eigendecomposition of the graph Laplacian matrix. Additionally, we propose a strategy to find optimal filters that are jointly  robust against the random graph  perturbations and  the signal noise.

\bibliographystyle{IEEEbib}
\bibliography{reference}



\end{document}